\documentclass[10pt,fleqn]{article}
\usepackage{amsmath}
\usepackage{amssymb} % e.g. \gtrsim
\usepackage{mathtools}
\usepackage{geometry}
\usepackage{newpxtext}
\usepackage{newpxmath}
\usepackage{dsfont} % Blackboard letters
\usepackage{caption} % font of captions is different from text
\usepackage{subcaption}
\usepackage{mathrsfs}	% e.g. nice Lagrange-L with \mathscr{L}
\usepackage{cite} % [1-5] instead of [1,2,3,4,5]
\usepackage{enumitem}% better control over list environments
\usepackage{array}% more options for table columns etc.
\usepackage{booktabs}% nicer tables
\usepackage{xurl}
\usepackage{cancel}
\usepackage{nicefrac}
\usepackage{pifont} % dinbgbats
\usepackage[linkcolor=blue,urlcolor=blue,citecolor=magenta,colorlinks=true]{hyperref}
\usepackage{cleveref}% clever cross referencing

\usepackage[nolist]{acronym}% acronym handling  

\usepackage{braket}
\usepackage{mleftright}\mleftright% better automatic sizes of parentheses
\usepackage{accents}
\usepackage{xfrac}
\usepackage{xspace}
\usepackage{soul}
\usepackage[textsize=tiny,backgroundcolor=yellow]{todonotes}
\makeatletter
\g@addto@macro\bfseries{\boldmath}% make section title math bold
\newcommand*{\defeq}{\mathchoice{\mathrel{\rlap{%
\raisebox{0.24ex}{$\m@th\cdot$}}%
\raisebox{-0.24ex}{$\m@th\cdot$}}%
=}{\mathrel{\rlap{%
\raisebox{0.24ex}{$\m@th\cdot$}}%
\raisebox{-0.24ex}{$\m@th\cdot$}}%
=}{\mathrel{\rlap{%
\raisebox{0.08ex}{\small$\m@th\cdot$}}%
\raisebox{-0.28ex}{\small$\m@th\cdot$}}%
=}{\mathrel{\rlap{%
\raisebox{0.08ex}{\tiny$\m@th\cdot$}}%
\raisebox{-0.28ex}{\tiny$\m@th\cdot$}}%
=}}
\newcommand*{\eqdef}{\mathchoice{=\mathrel{\rlap{%
\raisebox{0.24ex}{$\m@th\cdot$}}%
\raisebox{-0.24ex}{$\m@th\cdot$}}}{%
=\mathrel{\rlap{%
\raisebox{0.24ex}{$\m@th\cdot$}}%
\raisebox{-0.24ex}{$\m@th\cdot$}}}{%
=\mathrel{\rlap{%
\raisebox{0.08ex}{\small$\m@th\cdot$}}%
\raisebox{-0.28ex}{\small$\m@th\cdot$}}}{%
=\mathrel{\rlap{%
\raisebox{0.08ex}{\tiny$\m@th\cdot$}}%
\raisebox{-0.28ex}{\tiny$\m@th\cdot$}}}%
}
\makeatother
\allowdisplaybreaks
\newcommand{\ii}{\mathop{}\!\mathrm{i}\!\mathop{}}% imaginary i
\newcommand{\ee}{\mathrm{e}}% Euler e
\newcommand\simnsam{\mathrel{\ooalign{$\simeq$\cr
  \hidewidth\raise-.333ex\hbox{\rotatebox{45}{$\shortmid$}}\hidewidth\cr}}}
\DeclareMathOperator{\re}{Re}
\DeclareMathOperator{\im}{Im}

\newcommand{\ChargeC}{\ensuremath{\mathcal{C}}}

\newcommand{\ParityP}{\ensuremath{\mathcal{P}}}
\newcommand{\CP}{\ensuremath{\ChargeC\ParityP}\xspace}
\newcommand{\gCP}{\ensuremath{\widetilde{\ChargeC\ParityP}}\xspace}

\newcommand{\Z}[1]{\ensuremath{\mathds{Z}_{#1}}}
\newcommand{\U}[1]{\ensuremath{\mathrm{U}(#1)}}

\let\mathbb\mathds

\newcommand*\mytitle{Matter matters in moduli fixing and modular flavor symmetries}

\begin{document}
\begin{titlepage}
  \vspace*{1.0cm}
  
  \begin{flushright}
  UCI--TR--2023-03%
  \\
  \end{flushright}
  
  \vspace*{2cm}
  
  \begin{center}
  {\Large\sffamily\bfseries\mytitle}
  
  \vspace{1cm}
  
  \renewcommand*{\thefootnote}{\fnsymbol{footnote}}
  \begin{tabular}{>{\centering\bfseries\arraybackslash}m{0.8\linewidth}}
    V\'ictor Knapp--P\'erez$^a$, Xiang--Gan Liu$^a$, Hans Peter Nilles$^b$,\\ Sa\'ul Ramos--S\'anchez$^c$ and Michael Ratz$^a$
  \end{tabular}\\[8mm]
  \textit{$^a$\small
  ~Department of Physics and Astronomy, University of California, Irvine, CA 92697-4575 USA
  }
  \\[5mm]
  \textit{$^b$\small Bethe Center for Theoretical Physics and ~Physikalisches Institut der Universit\"at Bonn,\\ Nussallee 12, 
  53115 Bonn, Germany}
  \\[5mm]
  \textit{$^c$\small Instituto de F\'isica, Universidad Nacional Aut\'onoma de M\'exico, POB 20-364, Cd.Mx. 01000, M\'exico}
  \end{center}
  
  \vspace*{1cm}
  
  \begin{abstract}
    Modular flavor symmetries provide us with a very compelling approach to the flavor problem. It has been argued that moduli values close to some special values like $\tau=\ii$ or $\tau=\omega$ provide us with the best fits to data. 
We point out that the presence of  hidden ``matter'' fields, needed to uplift
symmetric AdS vacua, gives rise to a dynamical mechanism that leads to such values of $\tau$.  
   \end{abstract}
   \vspace*{1cm}
\end{titlepage}

\section{Introduction}

Modular flavor symmetries are an exciting approach to solve the flavor puzzles~\cite{Feruglio:2017spp}. 
Yukawa couplings get replaced by modular forms which depend on the so--called modulus $\tau$. 
The approach give rise to the phenomenon of ``Local Flavor Unification''~\cite{Baur:2019kwi,Baur:2019iai} 
with enhanced symmetries at the fixed points and lines of the modular flavor group.
This allows for flavor hierarchies of quark/lepton masses and mixing angles. 
In fact, it has recently been pointed out that values of the modulus $\tau$ close  to the fixed points 
$\ii$ and $\omega\defeq\exp(\nicefrac{2\pi \ii}{3})$ are favored by successful fits to the 
data~\cite{Feruglio:2021dte,Feruglio:2022kea,Feruglio:2023mii,Petcov:2022fjf,Abe:2023ilq}.
This seems to call for a dynamical mechanism that drives the modulus to the vicinity of the 
fixed points (and lines if we include \CP or a \CP--like symmetry). 

A particular appealing feature of modular flavor symmetries is its natural appearance in string theory. 
It has been known for some time that heterotic orbifold couplings are indeed modular forms~\cite{Lauer:1989ax,Lerche:1989cs,Chun:1989se,Lauer:1990tm}. 
This provides us with a consistent \ac{UV} completion of modular flavor symmetries as a starting point for top--down model building.
The central question to understand a dynamical mechanism that drives $\tau$ close to its fixed points can thus be addressed from the top--down perspective. 
This is, of course, nothing else but the widely discussed question of moduli stabilization in string theory.

This question about the localization of the modulus is the main focus of this paper. 
We suggest that such a mechanism should best be discussed in two separate steps:
\begin{enumerate}
\item finding a mechanism which in the case of unbroken symmetry favors 
solutions located exactly at, or close to, the fixed points,

\item identifying the dynamics of the theory that breaks the enhanced symmetry and
moves $\tau$ slightly away from the fixed loci.
\end{enumerate}
In what follows we shall outline a consistent scheme that gives rise to these two features. 

From the discussion of moduli stabilization of heterotic string theory, it is known that 
the scalar potential favors minima at the boundary of the fundamental domain~\cite{Cvetic:1991qm,deCarlos:1992kox}.
This even led to the conjecture of the absence of minima inside this domain~\cite{Cvetic:1991qm}.
Subsequent work, however, identified solutions inside the fundamental domain close to the fixed point $\tau=\omega$~\cite{Dent:2001cc,Dent:2001ut,Novichkov:2022wvg}.
Unfortunately, these solutions all have negative vacuum energy. This leads to \ac{AdS} space and is thus incompatible with the experimental observation that require \ac{dS} minima. 
In fact, there have been formulated various no--go theorems against the existence of \ac{dS} vacua in this context. 
An exhaustive discussion can be found in \cite{Leedom:2022zdm}.

Is this a severe problem? Probably not. 
The above results are based on models with very few fields. 
Similar no--go theorems for models with very few fields have been shown to no longer be valid in the presence of additional ``matter'' fields~\cite{Lebedev:2006qq}. 
In more realistic models, however, there are more fields that have to be included in the scalar potential. 
Some of them are needed, for example, to break the remnants of the traditional flavor symmetry that accompanies the modular flavor symmetry in the top--down approach. 
Typically, these fields also transform nontrivially under the modular  group~\cite{Nilles:2020tdp,Baur:2021mtl,Baur:2021bly}.
This leads to a picture reminiscent of an uplift via ``matter superpotentials''~\cite{Lebedev:2006qq,Lebedev:2006qc} in the discussion for the so--called \ac{KKLT} scenario.
This allows the realization of metastable \ac{dS} vacua as described by \ac{ISS}. 

Thus we arrive at a very satisfactory scenario. 
In a first step we obtain \ac{AdS} vacua at the boundary of the fundamental domain. 
The uplift mechanism in the second step shifts the modulus slightly away from the boundary giving rise to small parameters that can explain the hierarchies of masses and mixing angles in the quark and lepton sectors. 
The advocated two--stage scenario is thus naturally realized in this framework and gives a dynamical explanation for the existence of vacua in the vicinity of fixed loci of the modular group.

The paper is organized as follows. 
We proceed in \Cref{sec:ModulusStabilization} with a review of moduli stabilization in the presence of modular invariance, the appearance of \ac{AdS} vacua at the boundary of the fundamental domain as well as the few solutions within that domain that have been 
identified up to now. 
\Cref{sec:Uplift} will be devoted to the introduction of the uplift via ``matter superpotentials'' and spontaneous breakdown of modular invariance.
We describe this mechanism qualitatively within the approach of \ac{ISS} to obtain metastable \ac{dS} vacua. 
In \Cref{sec:Example} we present simple examples to illustrate our two--step process in which the spontaneous breakdown of modular invariance by matter fields and the uplift shifts the vacua away from the boundary of the fundamental domain.
In addition, we discuss the conditions on the uplifting scheme and the spontaneous breakdown of modular symmetry. 
Further discussion and outlook will be given in \Cref{sec:Outlook}.

\section{Modular invariant modulus stabilization}
\label{sec:ModulusStabilization}

Gaugino condensates~\cite{Nilles:1982ik,Derendinger:1985kk,Dine:1985rz} are standard non-perturbative ingredients in moduli stabilization. 
In torus--based compactifications these gaugino condensates respect target--space modular invariance~\cite{Font:1990nt,Nilles:1990jv,Dixon:1990pc,Mayr:1995rx}.
We will base our discussion on heterotic models and restrict the supergravity moduli sector to consist of the dilaton $S$ and a complex structure modulus $\tau$. 
Our (modular invariant) gaugino condensate is then described 
by~\cite{Font:1990nt,Nilles:1990jv,Cvetic:1991qm,Novichkov:2022wvg,Leedom:2022zdm}
\begin{equation}\label{eq:GeneralModularInvariantGauginoCondensate}
  \mathscr{W}_\text{gc}(S,\tau)=\frac{\Omega(S)\,H(\tau)}{\eta^6(\tau)}\;.
\end{equation}
Here, $\Omega(S)$ is typically chosen to be $B\, \ee^{-b\,S}$, where $B$ is a constant (see e.g.~\cite{deCarlos:1992kox}), and $b$ a $\beta$--function coefficient. 
In addition,
\begin{equation}\label{eq:ModularInvariantH}
  H(\tau)=\left(\frac{E_4(\tau)}{\eta^8(\tau)}\right)^n
  \left(\frac{E_6(\tau)}{\eta^{12}(\tau)}\right)^m
  P\bigl(j(\tau)\bigr)
  = \bigl(j(\tau)-1728\bigr)^{\nicefrac{m}{2}} \bigl(j(\tau)\bigr)^{\nicefrac{n}{3}}P\bigl(j(\tau)\bigr)\;,
\end{equation}
where the $E_k$ denote Eisenstein series, $j$ is Klein's modular invariant function, $P$ a polynomial thereof, and $n$ and $m$ are integers.

Many of the minima discussed in the literature occur at special points like $\tau=\ii$ , or other critical points, e.g.\ along the critical line $|\tau|=1$, on the boundary of  the fundamental domain of the extended modular group $\text{PGL}(2,\mathds{Z})\cong\text{PSL}(2,\mathds{Z})\rtimes \gCP$,
\begin{equation}
\label{eq:fundomain}
\mathcal{D}^*=\left\langle \tau \in \mathcal{H} ~\Big|~ -1/2\leq\re\tau\leq 0\;, \quad |\tau|\geq 1\, \right\rangle\;.
\end{equation}
This fact is easily understood from the general statement that minima often occur at symmetry--enhanced points~\cite{Michel:1971th,Feruglio:2019ybq}, and the observation that these values of $\tau$ lead to enhanced symmetries~\cite{Baur:2019kwi,Baur:2019iai}, as discussed in some detail in \Cref{app:enhancements}. 
In short, this symmetry enhancement arises from the invariance of certain points in $\tau$--moduli space under a finite set of $\text{PGL}(2,\mathds{Z})$ transformations, generated by the standard modular transformations
\begin{equation}\label{eq:tau_SandT}
 \tau\xmapsto{~S~}-1/\tau  \qquad\text{and}\qquad
 \tau\xmapsto{~T~}\tau+1\;,
\end{equation}
and the \CP or \CP--like transformation\footnote{For the distinction between physical \CP transformations and \CP--like transformations see~\cite{Chen:2014tpa}.} \cite{Dent:2001ut,Baur:2019kwi,Novichkov:2019sqv}
\begin{equation}\label{eq:tau_CP}
  \tau\xmapsto{~\gCP~}-\tau^*\;.  
\end{equation}
While it is true that extrema often prefer to occur at symmetry--enhanced points, 
they can also be found at locations that are not (obviously) special. In fact,
\cite{Dent:2001cc,Dent:2001ut,Kobayashi:2019xvz,Novichkov:2022wvg,Leedom:2022zdm} find minima of 
$\tau$ away from the critical line, i.e.\ away from the boundary of the fundamental domain. 
However, apart from the fact that these minima are \ac{AdS}, i.e.\ have negative vacuum energy, 
they are not continuously connected to the critical points in any known way.

In what follows, we focus on the question of how one may perturb the \ac{VEV} 
of $\tau$ to be parametrically close, but not identical, to special values such as $\ii$ or $\omega$. 
As we shall see, these corrections often also provide us with positive contributions 
to the vacuum energy, and may allow us to uplift unrealistic \ac{AdS} vacua.

\section{Uplifting and spontaneous breaking of modular invariance}
\label{sec:Uplift}

\subsection{Uplifting in flux vacua}

Negative vacuum energy is a problem that has been considered in connection with the so--called \ac{KKLT} scenario. \ac{KKLT} consider a modulus setting the value of the 4D gauge coupling, which we will denote $S$ in order to connect this discussion to \Cref{sec:ModulusStabilization}. 
$S$ is described by a superpotential and K\"ahler potential of the sort
\begin{subequations}\label{eq:toyKKLT}
\begin{align}
 \mathscr{W}_\text{toy KKLT}&=c-B\,\ee^{-b\,S}\;,\\
 K_\text{toy KKLT}&=-\ln(S+\overline{S})\;.  
\end{align} 
\end{subequations}
Here, $c$ is a constant that in the original \ac{KKLT} scenario gets induced by fluxes. 
In the framework of heterotic models, which we are primarily interested in, hierarchically small constants can emerge from approximate R symmetries \cite{Kappl:2008ie}. 
The scalar potential derived from \eqref{eq:toyKKLT} has an \ac{AdS} minimum at $S\sim \ln(c/Bb)$. \ac{KKLT} also have triggered  extensive research to identify possible solutions. 
In what follows, we will focus on \cite{Lebedev:2006qq}, where the uplift results from matter field interactions.

\subsection{Metastable dynamical supersymmetry breaking}
\label{sec:ISS}

Dynamical supersymmetry breaking \cite{Witten:1981nf} is an attractive scheme to explain why the scale of supersymmetry breaking, and hence the electroweak scale, is hierarchically smaller than the fundamental scale. 
While it is rather nontrivial to construct models without a supersymmetric ground state \cite{Shadmi:1999jy}, it has been noticed that rather straightforward models have metastable supersymmetry breaking vacua which are sufficiently long--lived \cite{Intriligator:2006dd}. 
In the vicinity of the metastable vacuum the model can be described by \cite[Section~2.2]{Intriligator:2007cp}
\begin{subequations}\label{eq:ISS_eff}
  \begin{align}
    K_\text{ISS,eff}&=X^\dagger X-\frac{(X^\dagger X)^2}{\Lambda_\text{ISS}^2}\;,\label{eq:K_ISS_eff}\\
    \mathscr{W}_\text{ISS,eff}&=f_X\,X\;.
  \end{align}
\end{subequations}
The second term in \eqref{eq:K_ISS_eff} may be thought of as locally describing the Coleman--Weinberg potential stabilizing the meson field $X$ at or close to the origin.

\subsection{Spontaneous breaking of modular invariance}
\label{sec:SSBofModularInvariance}

Modular invariance is nonlinearly realized and, in a sense, therefore spontaneously broken for 
a generic  \ac{VEV} of $\tau$. 
However, it can be broken additionally by \acp{VEV} of matter fields with  nontrivial modular weights. 
In explicit string models most of the matter fields have nontrivial modular weights. 
This is also true for those fields $\phi_i$ whose \acp{VEV} cancel the so--called \ac{FI} term $\xi_\mathrm{FI}>0$ in the $D$--term
\begin{equation}
\label{eq:FIterm}
  D = \xi_\text{FI} + \sum_i q^\mathrm{anom}_i |\langle \phi_i\rangle |^2\;,
\end{equation}
associated with a pseudo--anomalous $\U1_\mathrm{anom}$ gauge symmetry, where $q_i^\mathrm{anom}$ denote the $\U1_\mathrm{anom}$ charges of the fields with non--trivial \acp{VEV} $\langle \phi_i\rangle$.
In addition to modular flavor symmetries, string constructions exhibit ``traditional'' Abelian \cite{Binetruy:1994ru} and non--Abelian flavor symmetries~\cite{Kobayashi:2006wq,Olguin-Trejo:2018wpw}, which are defined as (flavor) symmetries under which $\tau$ transforms trivially.
These symmetries can combine with modular flavor symmetries in an eclectic scheme~\cite{Nilles:2020tdp,Nilles:2020gvu,Baur:2022hma} (also realized in bottom--up scenarios~\cite{Nilles:2020nnc,Chen:2021prl,Ding:2023ynd}), and can also be spontaneously broken by the \acp{VEV} of fields transforming nontrivially under them.

In earlier analyses it has been found that these field \acp{VEV} provide us with an expansion parameter of the order of the Cabibbo angle that breaks traditional flavor symmetries~\cite{Binetruy:1994ru}. 
Surveys of explicit string models such as~\cite{Lebedev:2008un} with the chiral matter content of the \ac{SM} under the \ac{SM} gauge group show that practically all models have an \ac{FI} term of the appropriate size, and \ac{SM} singlets which can acquire \acp{VEV} that cancel it.
This then leads to the interpretation of these \ac{SM} singlets as ``flavons'' of traditional symmetries like in the \ac{FN} model.
As we shall see below, the same \acp{VEV} lead to a small departure of the $\tau$ \ac{VEV} from the critical points or lines.

The alert reader may now be worried that, if modular invariance is broken by a number of fields, the predictive power of the scheme may disappear. 
This is, however, not the case in models which resemble top--down constructions, in which the modular weights of the couplings and fields have patterns which largely prevent us from replacing the modular forms by monomials of matter fields. 
This question will be studied in detail elsewhere. 
It is nonetheless worthwhile to recall that the predictive power of bottom--up models in this scheme is challenged by e.g.\ supersymmetry breaking effects~\cite{Criado:2018thu} and the  lack of control over the K\"ahler potential~\cite{Chen:2019ewa}.

\section{Examples}
\label{sec:Example}

In order to describe the ingredients outlined in \Cref{sec:ModulusStabilization}, 
we consider a system consisting of the dilaton $S$, the (complex structure) modulus 
$\tau$ and an \ac{ISS}--type matter field $X$.  
The K\"ahler potential and superpotential are given by

\begin{subequations}
\label{eq:ExampleKandW}
\begin{align}
 K&=-\ln(S+\overline{S})-3\ln(-\ii\tau+\ii\bar\tau)+(-\ii\tau+\ii\bar\tau)^{-k_X}\overline{X}X-(-\ii\tau+\ii\bar\tau)^{-2k_X}\frac{(\overline{X}X)^2}{\Lambda_\text{ISS}^2}\;,\\
\mathscr{W} &= \left(c_1 + c_2\,  \eta^{2k_c}(\tau) - B\,\ee^{-bS}+
 f_X\,\eta^{2(k_Y+k_X)}(\tau)X\right)\,\frac{H(\tau)}{\eta^6(\tau)}\;.
 \label{eq:ExampleW}
\end{align}  
\end{subequations}

In all the examples, $X$ settles at a very small yet nonzero value, consistently with the expectations and constraints outlined in \Cref{sec:ISS}. For simplicity, we set $P\bigl(j(\tau)\bigr)=1$ in the modular function $H(\tau)$.

\subsection{Fixing $\tau$ close to $\ii$} 
\label{sec:tau_close_to_i}

\begin{figure}[htb]
  \centering\includegraphics{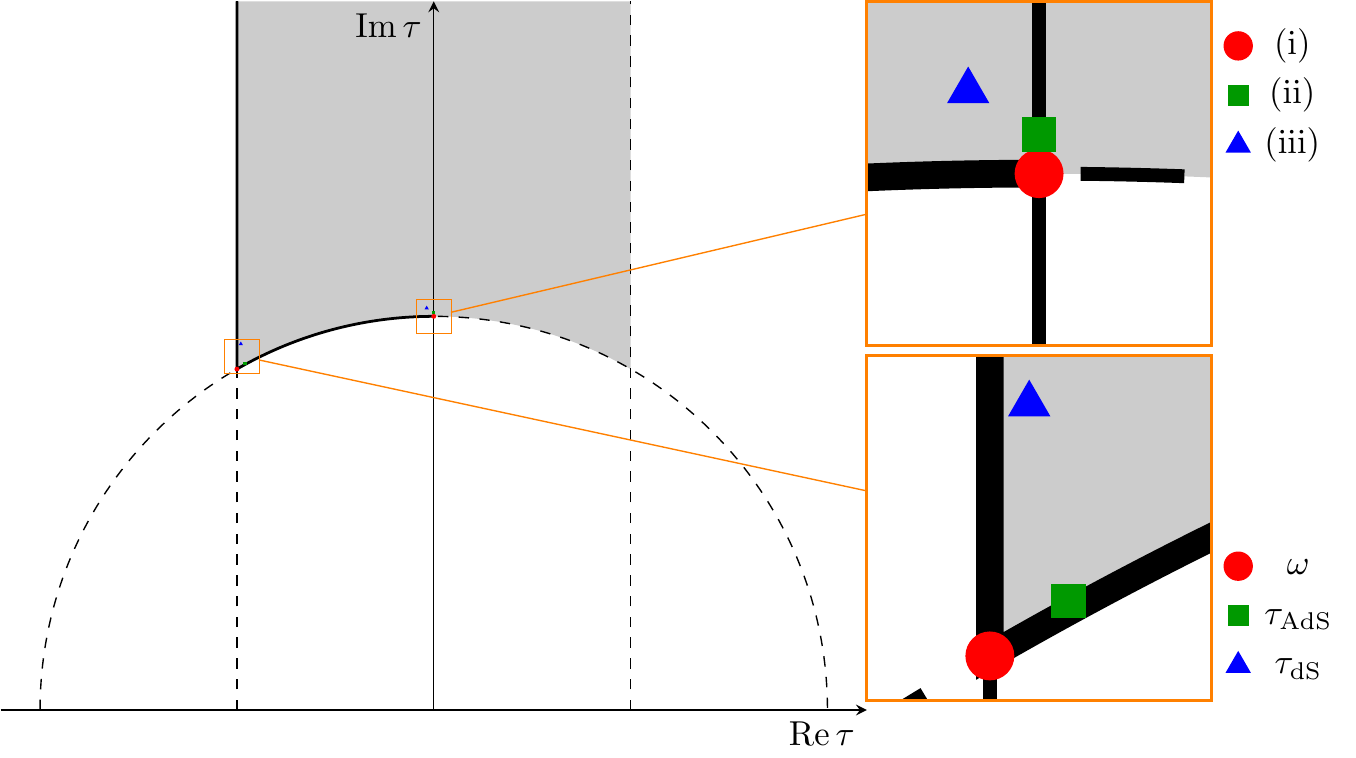} 
  \caption{Partial and full breakdown of modular invariance. 
  The upper inlay illustrates the stabilization of $\tau$ close to $\ii$ (cf.\ \Cref{sec:tau_close_to_i}) while the lower one refers to $\tau$ close to $\omega$ (cf.\ \Cref{sec:tau_close_to_omega}).}
  \label{fig:ModularTauVacua} 
 \end{figure}

In order to have realistic vacua in which $\tau$ is close to $\ii$, we set
\begin{equation}
 m=0\;,~ n= 1\;,~ c_1=2\cdot10^{-8}\;,~ k_c=1\;,~ k_X=0\;,~ b=10\;,~ B= 1~\text{and}~ \Lambda_\text{ISS}= 10^{-9}\;,
\end{equation}
where the dimensionful parameters are, as usual, given in Planck units.
It depends on the choices of the remaining parameters 
whether $\tau$ gets fixed
\begin{enumerate}[label=(\roman*)]
 \item precisely at $\ii$, leaving a residual $\mathds{Z}_2^{S}\times\mathds{Z}_2^{\gCP}$ (cf.\ \Cref{tab:listOfCriticalPoints}),
 \item along the imaginary axis with $\im\tau>1$, preserving a residual $\mathds{Z}_2^{\gCP}$, or 
 \item away from $\ii$ with $\re\tau\ne0$ and $\im\tau>1$.\vphantom{$\mathds{Z}_2^{\gCP}$} 
\end{enumerate}
We visualize these options in \Cref{fig:ModularTauVacua}, 
and provide examples in \Cref{tab:SampleParameters}. 

\begin{table}[htb]
 \centering 
 \begin{tabular}{*6{>{$}c<{$}}}
  \toprule 
  c_2 & f_X & k_Y & \tau & \langle\mathscr{V}\rangle \\
  \midrule 
  0 & 0 & 0 & \ii & <0 \\
  \ne0
  & 0 & 0 & -0.014
 +1.015\ii & <0 \\
  0 & 3.49\cdot10^{-8} & 0 & \ii & \simeq0 \\
  0 & 6\cdot10^{-8} & 1 & 1.010\ii & \simeq0 \\
  \ne0
  & 5\cdot10^{-8} & 0 & -0.018+1.011\ii & \simeq0 \\
  \ne0
  & 8.5\cdot10^{-8} & 1 & -0.018+1.021\ii & \simeq0 \\
  \bottomrule
 \end{tabular} 
 \caption{Survey of vacua close to $\ii$ with different residual symmetries. Our choice of nonzero values of $c_2$ is $c_2=2\ee^{\ii\pi/3}\cdot10^{-8}$, however our results are not qualitative impacted if we change the phase or magnitude of $c_2$, as long as the phase is nontrivial.
 The last column contains the vacuum energy, which is either negative ($<0$) or slightly positive ($\simeq0$).}
 \label{tab:SampleParameters}
\end{table}

As one can clearly see, in the absence of spontaneous breakdown of modular invariance as outlined in \Cref{sec:SSBofModularInvariance}, i.e.\ for $c_2$ and $k_Y=0$, $\tau$ remains at $\ii$.
However, once modular invariance gets broken spontaneously, $\tau$ somewhat deviates from $\ii$ but remains parametrically close to this special value.

\subsection{Fixing $\tau$ close to $\omega$}
\label{sec:tau_close_to_omega}

Let us next look at a model with parameters
\begin{equation}
m= 1\;,~ n= 0\;,~ c_1= 2\cdot 10^{-8}\;,~ c_2=0\;,~ 
b=10\;,~ B= 1\;,~ k_c=0\;,~f_X=0~\text{and}~\Lambda_\text{ISS}= 10^{-7}\;.
\end{equation}
As discussed in \cite{Dent:2001cc,Dent:2001ut,Novichkov:2022wvg}, this leads to an \ac{AdS} vacuum close to $\omega$. 
In more detail, the vacuum occurs at 
\begin{equation}
  \tau_{\text{\ac{AdS}}} \simeq -0.48+ 0.88\ii \;,~ 
  S\simeq 2.15~\text{and}~ \mathscr{V}\simeq -1.28\cdot 10^{-12}<0\;.
\end{equation}
Next we introduce an \ac{ISS} sector with 
\begin{equation}
\Lambda_\text{ISS}= 10^{-7}\;,
~ k_X= 0\;,~ 
k_Y=1 ~\text{and}~ f_X \simeq 6\cdot 10^{-8}\;.  
\end{equation}
This gives rise to a slight \ac{dS} vacuum at 
\begin{equation}
\tau_{\text{\ac{dS}}} \simeq -0.49+0.94\ii
~\text{and}~S\simeq2.16
\;.
\end{equation}
Note that $|\tau_{\text{\ac{dS}}}|\simeq1.06$ while $|\tau_{\text{\ac{AdS}}}|\simeq1.008$, i.e.\ uplifting moves $\tau$ further away from the boundary of the fundamental domain. This is illustrated in \Cref{fig:ModularTauVacua}.

Note that in all of these examples,  after the introduction of the uplift term that spontaneously breaks modular invariance,  the deviation $\delta\tau$ of the $\tau$ minimum from the critical points/lines is obviously continuously dependent on the continuous parameter $f_X$. 
Moreover, $f_X$ always falls within a certain interval, which is usually rather small: 
if $f_X$ was too small, the corresponding vacuum would be \ac{AdS}; 
on the other hand, if $f_X$ was too large, the minimum of $S$ would run away, i.e.\ $S\to\infty$. 
Therefore, $f_X$ is always limited to a certain range, which ensures that the deviation $\delta\tau$ from the critical points/lines remains nonzero yet parametrically small.

\section{Discussion and outlook}
\label{sec:Outlook}

We have revisited the question of moduli stabilization in the context of modular flavor symmetries. 
We point out that a number of no--go theorems which rule out realistic vacua no longer apply once matter field dynamics is taken into account. 
These matter fields provide us with two crucial ingredients: 
\begin{enumerate}
 \item spontaneous breakdown of modular invariance when fields of nontrivial modular weights attain \acp{VEV}, and 
 \item positive contributions to the vacuum energy from (metastable) dynamical supersymmetry breaking.  
\end{enumerate}
These ingredients are, in particular, realized in \ac{SM}--like string models in which \ac{FI} terms force certain matter fields acquire \acp{VEV}, and which are typically endowed with a hidden sector exhibiting nonperturbative strong dynamics at an intermediate scale. 

Historically, the \acp{VEV} induced by the \ac{FI} terms have been used in scenarios with traditional flavor symmetries, where the suppression of the \acp{VEV} against the fundamental scale was used to explain flavor hierarchies. 
In the context of the more recent models with modular flavor symmetries, it has been argued that data requires $\tau$ to be close to, but not exactly at, symmetry--enhanced points like $\tau=\ii$ or $\tau=\omega$. 
We find that the contributions from some ``hidden'' matter fields give us precisely that, namely small departures of $\tau$ from symmetry--enhanced points.  

This means, in particular, that certain string compactifications, which have been constructed to reproduce the (chiral) spectrum of the \ac{SM}, have all the ingredients to provide us with a successful theory of flavor.   
Altogether we are hence led to conclude that values of the modulus $\tau$ close to the special points $\tau=\ii$ or $\tau=\omega$ do have a clear top--down explanation.

\section*{Acknowledgments}

The work of V.K.-P., X.-G.L.\ and M.R.\ is supported by the National Science Foundation, under Grant No.\ PHY-1915005. 
The work of V.K.-P., S.R.-S.\ and M.R.\ is supported by C-MEXUS-CONACyT grant No.\ CN-20-38. 
The work of S.R.-S.\ is also supported by UNAM-PAPIIT IN113223, CONACYT grant CB-2017-2018/A1-S-13051, and Marcos Moshinsky Foundation.

\appendix
%%%%%%%%%%%%%%%%%%%%%%%%%%%%%%%%%%%%%%%%%%%%%%%%%%%%%%%%%%%%%%%%%%%%%%%%%%%%%%%%%%%%%%%%%%%%%%%%%%%%%%%%%%%
\section{Symmetry enhancement}
\label{app:enhancements}

If $\tau$ is located at a point or locus in moduli space that is left invariant under some discrete 
$\text{PGL}(2,\Z{})$ transformations, the ``traditional'' symmetries of the theory get enhanced.
Any of these symmetries is given by combinations of the $S$, $T$ and \gCP modular generators, 
of which the actions on $\tau$ are defined in~\eqref{eq:tau_SandT} and~\eqref{eq:tau_CP}. It is worth 
noting that in models based on modular flavor symmetries, the modulus vacua located at invariant 
points or loci tend to lead to trivial flavor mixing or unbroken  \CP  due to the presence of a
residual enhanced symmetry~\cite{Ding:2019gof,Ding:2021iqp}. As a result, more realistic modulus 
vacua typically need to deviate from the fixed points or loci. Interestingly, the fermion spectrum 
deviating from the fixed points exhibits a universal near-critical scaling behavior~\cite{Feruglio:2023mii}. 
As such, we often refer to these fixed points or loci as critical points or critical lines, emphasizing 
their potential association with critical phenomena.
A summary of the various critical points and lines along with their respective symmetry 
enhancements is presented in \Cref{tab:listOfCriticalPoints}.

Let us start by discussing the symmetry enhancements which occur at the three critical points, 
which coincide with the three vertices of $\mathcal{D}^*$ of~\eqref{eq:fundomain}. 
We see that $\tau=\ii$ is invariant both under the $S$--transformation in~\eqref{eq:tau_SandT} 
and the \CP--like transformation in~\eqref{eq:tau_CP}, i.e.\ it respects a $\Z2^S\times\Z2^{\gCP}$ symmetry.
On the other hand, $\tau = \omega \defeq \ee^{\nicefrac{2\pi\ii}{3}}$ is left invariant under $ST$,
\begin{equation}\label{eq:tau_ST}
\tau\xmapsto{~ST~} -1/ \left(\tau+1\right)\;,
\end{equation}
as well as the combined action of $T$ and~\eqref{eq:tau_CP}, leading to an enhancement by a $\Z3^{ST}\rtimes\Z2^{\gCP\circ T}$ symmetry.
The point at infinity, $\tau=\ii\infty$, is invariant under both the $T$ and the $\Z2^{\gCP}$ \CP--like transformations. 
Since the $\Z{}^T$ $T$--symmetry is an infinite discrete symmetry and does not commute with $\Z2^{\gCP}$, we obtain a 
$\Z{}^T\rtimes\Z2^{\gCP}$ symmetry enhancement.

The critical lines correspond to the boundaries of the fundamental domain $\mathcal{D}^*$ given in~\eqref{eq:fundomain}. 
Any point along the critical line $|\tau|=1$ is invariant under the $\Z2^{\gCP\circ S}$ combined action of $S$ and \gCP. 
Similarly, generic points along the line $\re\tau=\nicefrac{-1}{2}$ are invariant under the $\Z2^{\gCP\circ T}$ 
combined action of $T$ and \gCP. 
Finally, localizing the modulus along $\re\tau=0$ yields a $\Z2^{\gCP}$ symmetry enhancement.

\begin{table}[ht!]
\centering\small
\renewcommand{\arraystretch}{1.2}
\begin{tabular}{l*{6}c} \toprule
critical points/lines  & $\tau=\ii$ & $\tau=\omega$ & $\ii\infty$ & $|\tau|=1$ & $\re\tau=0$ & $\re\tau=\nicefrac{-1}{2}$  \\ \midrule
enhanced symmetries
                       & $\mathds{Z}_2^{S}\times\mathds{Z}_2^{\gCP}$ & $\mathds{Z}_3^{ST}\rtimes\mathds{Z}_2^{\gCP \circ T}$ & $\mathds{Z}^{T}\rtimes\mathds{Z}_2^{\gCP}$ & $\mathds{Z}_2^{\gCP\circ S}$ & $\mathds{Z}_2^{\gCP}$ & $\mathds{Z}_2^{\gCP \circ T}$ \\ \bottomrule
\end{tabular}
\caption{\label{tab:listOfCriticalPoints}
The critical points and lines in the fundamental domain of the (extended) modular group 
$\text{PGL}(2,\Z{})\cong\text{PSL}(2,\mathds{Z})\rtimes \gCP$, and the corresponding 
symmetry enhancements at those locations. More detailed information can be found e.g.\
in~\cite{Novichkov:2020eep,Baur:2021bly}.}
\end{table}

%%%%%%%%%%%%%%%%%%%%%%%% Bibliography
\providecommand{\href}[2]{#2}\begingroup\raggedright\endgroup

\begin{acronym}
 \acro{AdS}{anti de Sitter}
 \acro{dS}{de Sitter}
 \acro{EFT}{effective field theory}
 \acro{FN}{Froggatt--Nielsen \cite{Froggatt:1978nt}}
 \acro{FI}{Fayet--Iliopoulos \cite{Fayet:1974jb}}
 \acro{ISS}{Intriligator--Seiberg--Shih \cite{Intriligator:2006dd}}
 \acro{KKLT}{Kachru--Kallosh--Linde--Trivedi \cite{Kachru:2003aw}}
 \acro{SM}{standard model}
 \acro{SUGRA}{supergravity}
 \acro{SUSY}{supersymmetry}
 \acro{UV}{ultraviolet}
 \acro{VEV}{vacuum expectation value}
\end{acronym}

\end{document}